\documentclass[a4paper]{article}
\usepackage[T1]{fontenc}
\usepackage{mathptmx}
\usepackage{amsthm,amsmath,amssymb,enumerate}
\usepackage{graphicx}

\renewcommand\d{{\mathrm{d}}}
\newcommand\e{{\mathrm{e}}}
\newcommand\K{\,{\mathrm{K}}}
\newcommand\m{\,{\mathrm{m}}}
\newcommand\km{\,{\mathrm{km}}}
\newcommand\Hz{\,{\mathrm{Hz}}}
\newcommand\Mol{\,{\mathrm{Mol}}}

\newcommand\W{{\,\mathrm{W}}}

\title{Diurnal variation of VLF signals} \author{Richard
  Kaye\thanks{\texttt{rwkaye@gmail.com}.  This paper is copyright---please
contact me if you want to use any part of it for noncommercial uses.}}
\date{2014 Apil 10}

\begin{document}
\maketitle

\section{Introduction}

Like a number of amateurs, I have been recording VLF signal strength
using a home-made loop antenna, amplifier, and a computer sound-card.
Having obtained a number of days data, it seemed to be an interesting
exercise to fit the theoretically predicted $\log\sec$ variation with
zenith angle of the reflection layer height in the ionosphere
$D$-layer.  This short papers attempts just that.

I am no expert in radio astronomy, or radio engineering, or even of
the physics involved, though I do have some background knowledge and
expertise in mathematics and programming.  So much of this short paper
explains some of the theory I have learnt in the process of doing this
work.  I don't claim anything particularly new here, but some of the
techniques I use may be of interest and my explanations of the
background theory and description of the investigation here may be of
interest to other VLF amateurs.  There \textit{is} mathematics here,
at the upper end of the current A-level standard, including some
simple differential equations, but hopefully the presentation will be
straightforward enough for readers at this level.  It is quite
reassuring that some quite significant results on the ionosphere and
VLF needs nothing more complicated than this.

\section{The theory}

The $\log\sec$ variation of the height of the reflection layer is due
to Chapman~\cite{chapman}.  In the form needed for this work, the theory is
very straightforward and accessible to anyone with a knowledge of
simple differential equations.  I have learnt this theory from
reading Ratcliffe~\cite{ratcliffe} though no doubt many other texts are
available.  The following is a slightly simplified account that gives
the results needed.

The first stage (prior to applying Chapman's theory of the production
layer) is to understand the height variation of concentration of
particles (atoms, molecules, ions) in the atmosphere.

Let $h$ denote height (in $\m$) above some reference level (for
convenience, the Earth's surface) and $n=n(h)$ the concentration (in
$\Mol \m^{-3}$) of some species of molecule relevant to a particular
ionisation process, such as NO. If each molecule has mass $m$ and $g =
g(h)$ is the acceleration due to gravity then the force downwards
due to gravity on the molecules in the unit volume is
$nmg$.  This is balanced by the difference in pressure $p$
between the top and bottom of the volume so that
\begin{equation}
\frac{\d p}{\d h} = -nmg.
\end{equation}
Pressure is given by $p = nkT$ where $k$ is Boltzmann's constant and $T
= T(h)$ is temperature in $\K$, so
\begin{equation}
\frac{\d}{\d h}\left(nkT\right) = -nmg.
\end{equation}
Now for the range of heights to be taken here that are relevant to the
lower ionosphere, $h$ ranging from around 60 to 95 $\km$, and compared
to the radius of the Earth of $6371\km$, both $g$ and $T$ may be assumed
to be more or less constant.  Thus
\begin{equation}
\frac1n \frac{\d n}{\d h} = -\frac{mg}{kT}.
\end{equation}
We set $H = kT/mg$ (ideally with values of $g$ and $T$ at or near
those at the D-layer) and call $H$ the \textit{scale height} or
\textit{distribution height} of the species represented by $n$.
The solution of the differential equation above is
\begin{equation}
n = n_0 \e^{-h/H} \label{eq:n}
\end{equation}
where $n_0$ is a constant representing the value of $n$ at the 
reference height $h=0$.
In other words, concentration $n$ is theoretically an inverse exponential distribution 
with constant $H$.  

For readers unfamiliar with this and unclear on the significance of
$H$, this distribution is somewhat similar to the familiar
distribution in time of the number $N_02^{-t/t_0}$ of atoms of a
radioactive element undergoing decay.  The constant $t_0$ (a
length of time) is the \textit{half-life} of the the element, and waiting one
half-life results in halving the number of atoms.  Similarly, the
scale height $H$ is the height one must travel upwards to decrease $n$
by a factor of $\e = 2.71828\ldots$.  For the D-layer it is typically 
about $5\km$, as we shall see.

The next stage is to imagine ionising radiation being applied from
above, i.e.~from the sun.  The sun, we shall assume, is at an angle
$\chi$ from the zenith, i.e.~$\chi=0$ corresponds to the sun being
directly overhead and $\chi=90^\circ$ it being on the horizon.  Just
as it was the case that not all air molecules are relevant for
ionisation of the D-layer, so it is that not all frequencies are
relevant here either.  We will assume that a band of frequencies are
responsible for ionisation, and the power flux from the sun in this
band is $I_\infty$, measured in $\W \m^{-2}$, so if an area of one square
metre were mapped out in space on a plane perpendicular to solar rays, in one
second $I_\infty$ Joules of energy in the relevant band would pass though this
area.  As should be clear, if the plane were not perpendicular to the
solar rays the effective area available is less and less energy would
pass though it.  In fact if the solar radiation were at an angle
$\chi$ to the plane's perpendicular then $I_\infty\sec\chi$ Joules of energy would pass through the
plane, where $\sec\chi = 1/\cos\chi$, and again $\chi=0$ refers to
the rays being exactly perpendicular i.e.~directly above.

The sun's energy is absorbed by the atmosphere, and the amount it is
absorbed by is proportional to $n$---the constant of proportionality
(called the `absorption cross-section') will be denoted $\sigma$.  So
the ionisation radiation $I$ varies with height $h$ and as it passes through each unit
of volume is decreased by $\sigma n I\sec\chi$.  Thus the differential equation for $I$ is
\begin{equation}
\frac{\d I}{\d h} = \sigma n I\sec\chi
\end{equation}
(Some care is needed to get the sign right here, but the above is
correct since the ionisation energy is coming from above and is
absorbed in the atmosphere, so $I$ is decreasing as $h$ decreases.)
Our previous equation \eqref{eq:n} can be substituted in here and the
equation rearranged to give,
\begin{equation}
\frac 1I \frac{\d I}{\d h} = \sigma n_0 \sec\chi\, \e^{-h/H}
\end{equation}
which when solved gives
\begin{equation}
\log (I / I_\infty) = -H \sigma n_0 \sec\chi\, \e^{-h/H},
\end{equation}
$\log$ being natural logarithm to base $\e$, or
\begin{equation}
I  = I_\infty \exp( -H \sigma n_0 \sec\chi\, \e^{-h/H} ).
\end{equation}

The energy absorbed by the atmosphere doesn't disappear but goes somewhere: it is either
converted to heat or used to ionise the atmosphere.  Thus the
production rate $q$ of electrons (or other charged particles that can
reflect radio waves) is proportional to the amount of energy
$\sigma n I\sec\chi$ absorbed. Letting $C$ denote the constant of
proportionality, we have
\begin{equation}
q = C \sigma n I\sec\chi
\end{equation}
or
\begin{equation}
q = C \sigma n_0 \e^{-h/H} \sec\chi \, I_\infty \exp( -H \sigma n_0 \sec\chi\, \e^{-h/H} ).
\end{equation}
To complete the story, the electrons produced in this way either
diffuse to a different height or recombine with other molecules in the
air according to one of a number of possible reactions.  More details on this are not
needed here.  What we need to observe here is that the height $h_m$ of the reflecting
layer corresponds to the position of greatest rate of electron production. (The 
reason why this is the right condition is slightly complicated,
but my understanding is that it is because the
rate of electron recombination is proportional to the concentration of electrons
and to the concentration of particles with which they can combine with.  It is these
simple proportionalities that ensure that the place of greatest change of electron 
concentration is the same as the place of greatest electron production.)
Thus to find the height $h_m$ of the reflecting layer we find the height
where $q$ is maximum, and the simple technique of differentiating
$q$ and setting the derivative equal to zero is used.  The derivative
of $q$ is obtained by the chain and product rules (noting the double
exponential in $h$) and $\d q/\d h = 0$ simplifies to
\begin{equation}
-\frac 1H + \sigma n_0 \sec\chi\, \e^{-h_m/H} = 0
\end{equation}
or 
\begin{equation}
h_m = H\log( H\sigma n_0 \sec\chi ) = H\log( H\sigma n_0) + H\log \sec\chi 
\end{equation}
which is the equation alluded to in the introduction.
Notice that this is of the form $h_m = A + H\log \sec\chi$ where
$H$ is the scale height, a value of some physical importance.
The constant $A = H\log( H\sigma n_0)$ represents the height the
reflecting layer would have been at, given steady conditions with the same 
radiation energy but with the sun exactly overhead.

Of course, a VLF receiver does not measure the height of the
reflecting layer directly, but this height can sometimes be inferred from
measurements. The varying strength of a signal is an indication of
an interference between different paths of propagation.  
Normally, there are many different paths, but for signals from 
nearby transmitters we can reasonably model the process 
as being the interference effects between a ground-wave and a bounced 
sky-wave. This part of the modelling 
process is essentially just one of geometry without any calculus.
The details have no doubt been worked out many times, I summarise the
results here, and Mark Edwards~\cite{edwards} gives more detail
and additional explanations should they be required. 

Given that the ground-wave travels a distance $D$ along the curved
surface of the earth and the sky-wave travels a distance $L$, the phase
difference between them (in radians) is
\begin{equation}
\phi = 2\pi(L-D)f/c + \pi
\end{equation}
where $f$ is the frequency of the transmission (in $\Hz$), $c$ is
the speed of light and the additional $\pi$ is due to a phase change
on reflection.  By geometry and the cosine rule, the distance $L$ is 
related to $D$, $h_m$ and $R$ (the
radius of the Earth) by
\begin{equation}
L= 2\sqrt{ R^2 + (h_m+R)^2 - 2R(h_m+R)\cos(D/2R) } 
\end{equation}
and the power $P$ of the received wave is proportional to
\begin{equation}
G^2 + S^2 +2GS\cos\phi
\label{eq:measureintermsofphi}
\end{equation}
where $G$ is the amplitude of the ground wave and $S$ the amplitude of
the sky-wave.  

\section{The practice}

The proposal is to look at the variation of the received power over
the course of a quiet day, for a nearby transmitter and see how well
the observed data for the theoretical pattern described here.  Note
that there are four unknown variables in the theory: the scale height
$H$, the quantity $H\log( H\sigma n_0)$ representing the height of the
reflecting layer at $\chi = 0$, and the amplitudes of the ground and
sky-waves. 

Thus we want to fit 
\begin{eqnarray}
 \text{power} &=& Q + 2P\cos\phi \label{eq:cosphi}\\
 \phi &=& 2\pi(L-D)f/c + \pi \\
 L &=& 2\sqrt{ R^2 + (h_m+R)^2 - 2R(h_m+R)\cos(D/2R) } \\
 h_m &=& A + H\log\sec\chi \label{eq:chapman}\\
 \chi &=& \text{sun's zenith angle at midpoint}
\end{eqnarray}
to our data, where $Q=G^2+S^2$ and $P=GS$ in \eqref{eq:measureintermsofphi}. 
I used a downloadable
algorithm for the sun's zenith angle\footnote{From
  \texttt{http://www.psa.es/sdg/sunpos.htm}} and readily available
data on the position
of the transmitter---and hence derived the longitude and latitude of
the midpoint. Thus the unknowns are $Q,P,A,H$ only.  

In any curve fitting algorithm, having initial estimates for the
unknown values being sought is very useful indeed. In this case the 
constants can be given rough
estimates quite quickly: $H$ is known to be about $8\km$ at ground
level, independent of the species involved~\cite[page 5]{ratcliffe};
the height of the reflecting layer is nominally around $90\km$; and
the quantities $Q = G^2+S^2$ and $2P = 4GS$ can be estimated quickly from the
VLF measurements, as follows.  In the daytime, excluding
some complicated sunrise/sunset effects which are due to
more complicated geometry of a spherical Earth and different 
propagation paths, the measured signal varies from a
minimum at $\phi = 2N\pi-\pi/2$ to a maximum at $\phi = 2N\pi+\pi/2$
(for some integer $N$ which cannot be directly estimated) 
and thus from \eqref{eq:measureintermsofphi} the
difference between this maximum and minimum is about $4GS$.
The quantity $G^2 + S^2$ is then the value exactly halfway between 
this maximum and minimum.  

Even with these initial guesses for the parameters involved, I do not
have an ideal curve-fitting algorithm.  The main problem is that
rather different height estimates sometimes record a `good fit' simply
because the $\cos\phi$ function in \eqref{eq:cosphi} is periodic and
differing values of $\phi$ do indeed give reasonably good fits.  To
say this in another way, it is not really possible to obtain the
height $h_m$ from the measured phase information as the mapping 
from $h_m$ to $\phi$ is many-to-one.  A second problem is the possible
occurrence of SIDs---periods when the data do not fit the usual quiet
diurnal pattern.

As a compromise, my experimental algorithm discounts a certain percentage of the
data (say 10\%, though this parameter can be varied).  The measure of
`fit' is the sum of the $(v_{\text{measured}}-
v_{\text{predicted}})^2$ for all but the 10\% greatest values of this
quantity. (These squared differences are stored in a heap so that the
best 90\% can be extracted quickly.)  Rather than risking a `clever'
algorithm rapidly settling on a `bad' value of $h_m$, I test many
values of $A,H$ differing by only a small amount in succession before
selecting the `best' and then refining this value in a similar way.
But as it turns out, the curve fitting is relatively stable in the
other two parameters $S,P$ so that it is possible to find reasonably
good values for $A,H$ using the initial estimates for $S,P$, using
these values to refine the estimates for $S,P$, and then using these
values to refine the values for $A,H$.  Exactly how often this process
should continue and in what order and with what step size is still 
very much open for
experiment, but as can be seen, reasonably good fits can indeed be obtained.

In the month of October 2013, the 21st was a comparatively `quiet' day
and will be used to illustrate these methods.  The signal  from Skelton, UK,
on 22.1kHz is the strongest nearby signal at my location, being about 263km distant.
I entered the coordinates of the midpoint and the distance to the transmitter
and started to fit the data.  This was the result.

\begin{center}
\resizebox{\textwidth}{!}{\includegraphics{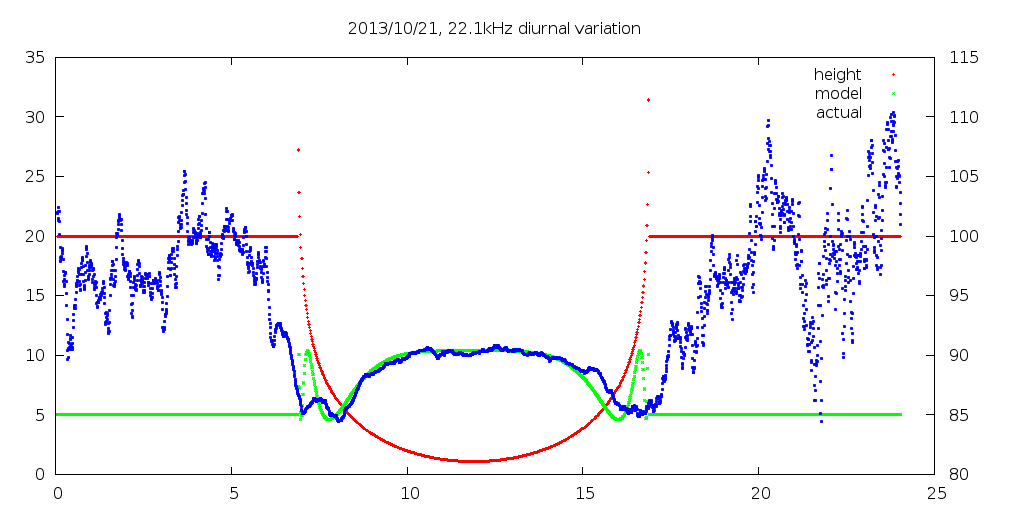}}
\end{center}

The blue line shows the actual measured values. The red line shows the 
model's value for the height of the reflecting layer---which is
only defined for $0\le \chi <90^\circ$ since $\sec\chi$
approaches infinity near sunset. (Outside this region I arbitrarily
set it to 100km.) The green line
shows the fitted curve.

The fit seems reasonably good, though not by any means perfect.
The values the fit took for the height of the reflecting layer at 
various times and for $A,H$ in \eqref{eq:chapman} above were
$A=76.84$ and $H=5.06$.
These values seem encouraging, especially as they were chosen from 
the fitting algorithm over a range of plus or minus 20\% and we read
\begin{quotation}
For VLF waves incident
on the ionosphere at steep incidence, the reflection height, $h$, appears to vary as
$h_0 + H\ln\sec\chi$
where $\chi$ is the solar zenith angle. $h_0$ is about 72 km, and $H$ is about 5 km, which
happens to be the scale height of the neutral gas in the mesosphere.
\end{quotation}
(from Hunsucker and J. K. Hargreaves~\cite[page 35]{hunsuckerandhargreaves}).

Unfortunately, one worry is that (as already mentioned) quite
different values for the height parameter also fit quite well through
using a different period in the $\cos \phi$ function.  
For example, the following fit
\begin{center}
\resizebox{\textwidth}{!}{\includegraphics{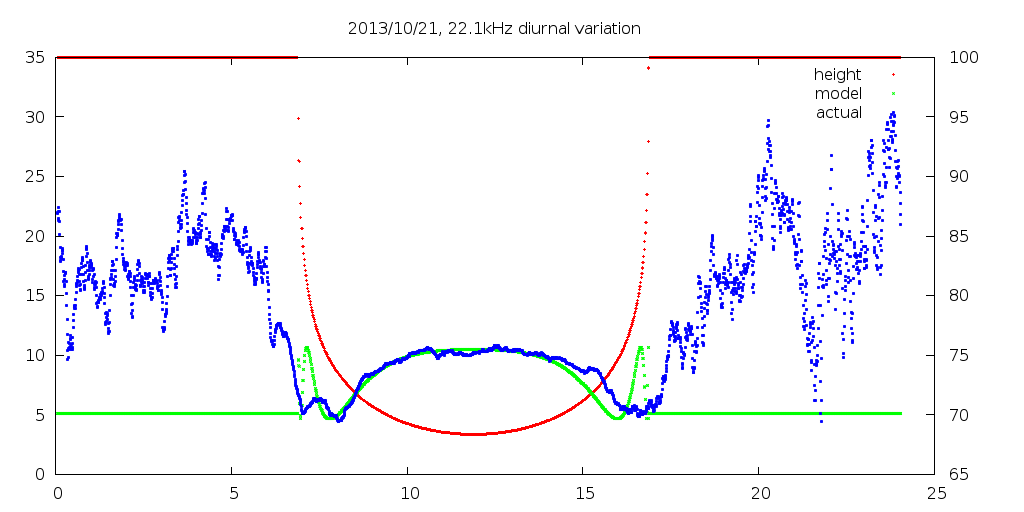}}
\end{center}
was found with parameters $A=64.06$ and $H=5.12$ showing that value
obtained by the fit for $A$ is not particularly robust.  Similarly
values for $H$ from reasonable looking fits were found ranging from
$5$ to $6$.

One possible approach is to look at a number of different signals and compare them.
For example, this is the nearby Anthorn signal on 19.6kHz on the same day.
\begin{center}
\resizebox{\textwidth}{!}{\includegraphics{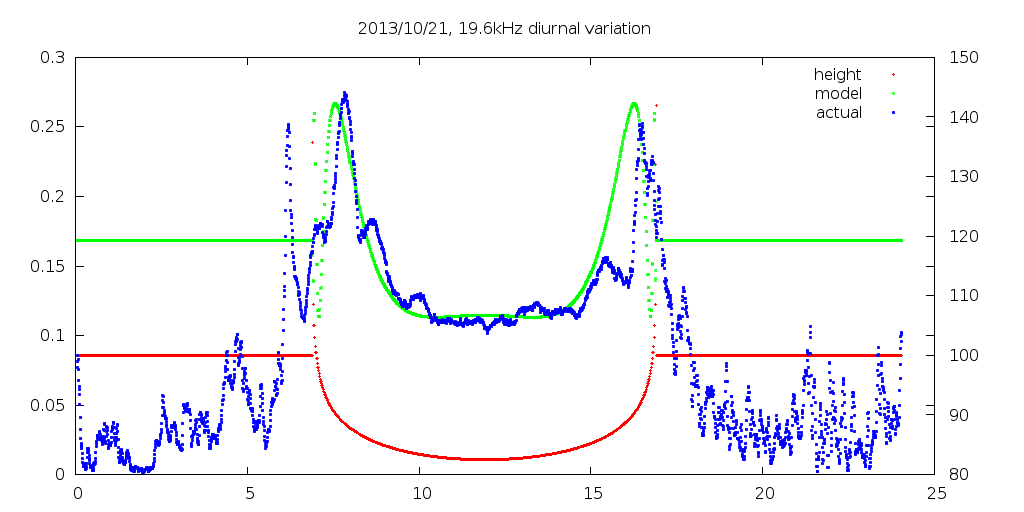}}
\end{center}
The fit here had $A=78.23$ and $H=5.06$.  This suggests that
these parameters are in the right sort of `ballpark', but the evidence isn't
particularly convincing.

Mark Edwards has pointed out (especially in his presentation to the
BAA Radio group in 2011) that the \textit{combination} of these two
signals from two transmitters very close together can
\textit{together} give an accurate fix on the height of the reflecting
D-layer, because they are operating at different frequencies and it so
happens that at his location the result is that the daytime signals
from these two locations appear almost a mirror image of each other.
What's more, the reflection points for the sky-waves for these two
transmissions are very close to each other so it is reasonable to
assume that the D-layer height is the same in both cases. In this
context the simultaneous fit of the log sec model to these data may be
only feasible for $A$ around $77$ or $78$ and $H$ about $5$.  This is
a very sensible suggestion and well worth undertaking where feasible,
but in general this will depend on specific local circumstances (such
as the availability of suitable nearby transmissions and the distance
to the transmitters and frequency of these transmissions).  In
general, the hope is that an intelligent examination of all the
various possible heights in the case of two or more separate
transmissions will rule out all but the correct D-layer height,
especially if the reflection points in question are very close
together.  There is obviously more work to be done here.

In both cases, the fit is noticeably not so good near sunrise and
sunset, particularly near sunset.  Of course one cannot expect a perfect fit near
these limits, because for example the model predicts an infinite
height at sunrise/sunset, whereas in fact the curvature of the Earth
has effects that are not taken into account by the model (such as
the possibility that, at 90km above the ground, the ionosphere is
radiated by solar radiation even when $\chi$ is greater than $90^\circ$).
Also, other propagation paths come into effect at such extremes, and
other mechanisms ionising mechanisms
(such as cosmic rays) will become more significant at such times.
Some indications that different mechanisms are at play are already
evident in the multiple peak structure in the sunrise/sunset pattern for
19.6kHz, which (since the peaks are not at the maximum) cannot be
predicted by the simple Chapman model with a single ionising source,
and perhaps suggests evidence for more than one source of ionisation.
This could be investigated further.  Possible improvements to the
model include: (a)~reworking it for a spherical Earth;
(b)~incorporating any tilt of the D-layer into the calculations, since there
is no particular reason why the D-layer will always be horizontal, especially at
sunset and sunrise; and (c)~investigating other ray paths.
For~(c), Edwards~\cite{edwards2} reports improvements when an additional
double bounce model is added.

Irrespective of the situation at sunrise, the shape of the measured
and modelled curves are rather different at sunset,
though a casual look at the data prior to making these attempts at
fitting the model to them did not suggest there might be a problem.  A
little investigation explains why.

The next graphic shows the same raw data alongside GOES satellite
measurements of X-ray solar flux.

\begin{center}
\resizebox{\textwidth}{!}{\includegraphics{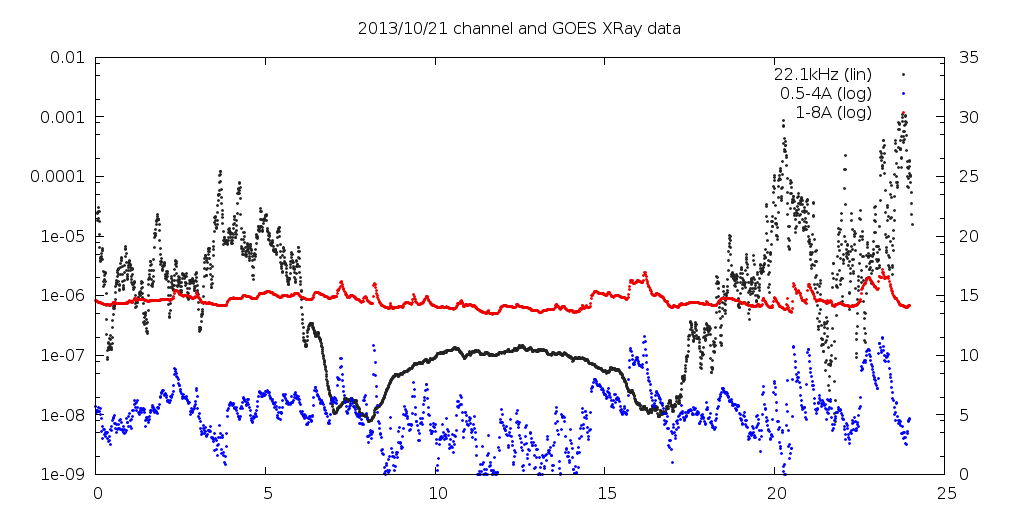}}
\end{center}

One sees there was enhanced solar X-ray activity from 15:00UT onwards,
and particularly from 15:40UT. At it maximum (at 16:12UT) this was at
the C2.7 level, which is often small enough to be neglected, and in
this case not sharp enough to be an obvious `flare' creating a peak in
the VLF trace. It seems highly likely that the lack of `fit' at this
time and the enhanced solar X-ray activity are related.  Indeed this
seems to the main value for this technique: that comparing measured
data with the model, the places where the measured data does not fit
are more obvious and these often will reflect some interesting
phenomena going on---in this case a minor X-ray induced ionospheric
disturbance---that might have been easy to miss otherwise.  Or to put
it another way, such analyses have the potential to dramatically
enhance the sensitivity of the measurements without changing the
hardware in any way.

\section{Conclusions}

Fitting the Chapman model of diurnal variation can be done, and
often seems successful except very close to the points of sunrise and sunset
where the model (at least in the form given here) 
is not meaningful.  However drawing conclusions from
the model fitting has its difficulties, mainly because the
mathematical mapping of reflection layer height (as predicted by the
model) to phase difference (as measured) is many-to-one, hence
different heights can results in fits that are or appear to be just as
good.  Any further experiments that exploit this model to obtain
measurements of (for example) the scale height will have to resolve
this problem and make very clear why the values for heights chosen are
indeed the correct ones.

Nevertheless, even if the actual numerical values obtained from the process
are not believed, the technique
can provide a source of evidence for ionospheric disturbances measured from
VLF data near sunrise or sunset when no obvious traditional `SID pattern' is
present in the data.  In other words, these techniques can in principle be
used to dramatically increase the sensitivity of a SID detector especially
near sunrise/sunset.

\nocite{*}
\bibliographystyle{plain}
\bibliography{vlf}
\end{document}